# Industrial Practices of Requirements Engineering for ML-Enabled Systems in Brazil

Antonio Pedro Santos Alves
apsalves@inf.puc-rio.br
Pontifical Catholic University of Rio
de Janeiro (PUC-Rio)
Brazil

Marcos Kalinowski
kalinowski@inf.puc-rio.br
Pontifical Catholic University of Rio
de Janeiro (PUC-Rio)
Brazil

Daniel Mendez
daniel.mendez@bth.se
Blekinge Institute of Technology
(BTH)
Sweden

Hugo Villamizar
hvillamizar@inf.puc-rio.br
Pontifical Catholic University of Rio
de Janeiro (PUC-Rio)
Brazil

Kelly Azevedo
kellyableal@gmail.com
Pontifical Catholic University of Rio
de Janeiro (PUC-Rio)
Brazil

Tatiana Escovedo
tatiana.inf.puc-rio.br
Pontifical Catholic University of Rio
de Janeiro (PUC-Rio)
Brazil

Helio Lopes
lopes@inf.puc-rio.br
Pontifical Catholic University of Rio
de Janeiro (PUC-Rio)
Brazil

## ABSTRACT

**[Context]** In Brazil, 41% of companies use machine learning (ML) to some extent. However, several challenges have been reported when engineering ML-enabled systems, including unrealistic customer expectations and vagueness in ML problem specifications. Literature suggests that Requirements Engineering (RE) practices and tools may help to alleviate these issues, yet there is insufficient understanding of RE's practical application and its perception among practitioners. **[Goal]** This study aims to investigate the application of RE in developing ML-enabled systems in Brazil, creating an overview of current practices, perceptions, and problems in the Brazilian industry. **[Method]** To this end, we extracted and analyzed data from an international survey focused on ML-enabled systems, concentrating specifically on responses from practitioners based in Brazil. We analyzed the cluster of RE-related answers gathered from 72 practitioners involved in data-driven projects. We conducted quantitative statistical analyses on contemporary practices using bootstrapping with confidence intervals and qualitative studies on the reported problems involving open and axial coding procedures. **[Results]** Our findings highlight distinct aspects of RE implementation in ML projects in Brazil. For instance, (i) RE-related tasks are predominantly conducted by data scientists; (ii) the most common techniques for eliciting requirements are interviews and workshop meetings; (iii) there is a prevalence of interactive notebooks in requirements documentation; (iv) practitioners report problems that include a poor understanding of the problem to solve and the business domain, low customer engagement, and difficulties managing stakeholders expectations. **[Conclusion]** These results provide an understanding of RE-related practices in the Brazilian ML industry, helping to guide research and initiatives toward improving the maturity of RE for ML-enabled systems.

## 1 INTRODUCTION

Machine Learning (ML) has increasingly gained prominence in the global industry. In 2023, 41% of Brazilian companies have already implemented ML to some extent, while the remaining are willing to adopt it as well [16]. These systems, where ML components are integral parts of larger systems, are known as ML-enabled systems. Their behavior is based on explicitly defined rules and data used by the ML component to make predictions [40].

Transitioning from traditional software to ML-enabled systems poses various challenges from the viewpoint of Software Engineering (SE) [33]. Some examples include covering additional quality properties such as fairness and explainability, dealing with a high degree of iterative experimentation, and mismatched assumptions in customers and multidisciplinary teams [29, 36]. Such challenges typically demand extra effort to successfully develop ML-enabled systems and may contribute to the statistic that 87% of ML projects never reach production [13].

Due to the communication and collaboration-intensive nature, as well as inherent interaction with most other development processes, the literature suggests that Requirements Engineering (RE) can help address several challenges when engineering ML-enabled systems [1, 42, 44]. However, establishing effective RE practices in ML projects may be difficult primarily due to (i) the lack of practitioners engaged in formal RE activities [3], and (ii) the absence of tailored techniques and tools for data-driven projects since research on this intersection mainly focuses on using ML techniques to support RE activities rather than exploring how RE can improve the development of ML-enabled systems [8]. Therefore, it is not surprising that recent studies emphasize that practitioners find RE as the most difficult activity of ML projects [18, 27, 35].

In order to strengthen the empirical evidence into current Brazilian industrial RE practices, perceptions, and challenges when developing ML-enabled systems, we extracted and analyzed data from an international survey focused on current practices and challenges for ML-enabled systems, concentrating specifically on responses from practitioners based in Brazil. We looked deeper into the RE-related answers gathered from 72 practitioners involved in ML projects in Brazil afterward. Based on practitioners' responses, we conducted quantitative and qualitative analyses, providing insights into (i) what role is typically in charge of requirements; (ii) how requirements are typically elicited and documented; (iii) which non-functional requirements typically play a major role; (iv) which RE activities are perceived as most difficult, and (v) what RE-related challenges do ML practitioners face. We share our findings on the practices, perceptions, and challenges of RE for ML in Brazil to contribute to a broader understanding of the field. By analyzing the data from this specific region, we aim to provide valuable context that could guide future research and compare trends in RE for ML on a global scale.

The remainder of this paper is organized as follows. Section II provides the background and related work. Section III describes the research method. Section IV presents the results. Sections V and VI discuss the results and threats to validity. Finally, Section VII presents our concluding remarks.

## 2 BACKGROUND AND PREVIOUS WORK

Machine Learning (ML) is a sub-field of artificial intelligence that involves the study of algorithms and statistical models that allow software systems to learn and make predictions based on data [19]. By recognizing patterns in the data on which they are trained, ML algorithms are developed to improve automatically over time on unseen data [34]. Consequently, the development of ML-enabled systems differs significantly from conventional software systems due to several key factors.

There is a high level of experimentation and uncertain outcomes when developing ML-enabled systems [2] and a multidisciplinary team is essential, comprising domain experts, software developers, data science, and engineering professionals [36]. Data scientists, who typically take the rein when developing ML projects [26], experiment with various data, algorithms, and models to determine the most effective approach for achieving their objectives, which means that setting up goals and requirements at the beginning of the process would demand an estimate of different metrics (*e.g.*, accuracy) in advance [18].

Uncertainty and experimentation are expected for this scenario once ML projects often begin as small Proof-of-Concept (PoC) initiatives, and 87% of them never reach production [13]. The complexity of transitioning from laboratory-level models to production architectures brings several challenges [30, 48]. Although ML-enabled systems are hugely popular and in demand, multiple ML projects that have overcome the first barrier of reaching production have failed in recent years, leading to severe repercussions for the organizations involved and the society at large [5, 12]. The reason for this is often the same: systems that incorporate ML components tend to put stakeholder needs in the background and oversimplify important scenarios and trade-offs. This leads to a problem that the Requirements Engineering (RE) discipline can tackle.

Requirements Engineering constitutes approaches to understanding the problem space and specifies requirements that all stakeholders agree upon [9]. As such, it concentrates on understanding the actual problem, what is needed towards a system result, and how to resolve potential conflicts, and it is thus characterized by the involvement of interdisciplinary stakeholders and often resulting in uncertainty [45]. The large degree of uncertainty in developing ML-enabled systems introduces new challenges and heavily affects RE [6, 33].

In order to overcome such difficulties, some studies have proposed new methods or adapted existing ones to handle requirements on such systems [17, 43]. However, gathering empirical evidence from the industry is essential to accurately identify real-world challenges, perceptions, and current practices. For instance, several studies have surveyed practitioners and found that unpredictability makes it difficult to define any criteria or requirements regarding the output of ML components [3, 7, 44]. This introduces a challenge in collaboration with stakeholders, who may perceive what ML is capable of wrongly [14].

We advocate that insights from practitioners can guide the development of new RE techniques for ML, thereby increasing the likelihood of designing and developing ML-enabled systems that meet customer needs and potentially avoid costly problems later on. To complement the already discussed research, we present additional empirical evidence from Brazil on the current practices, perceptions, and challenges regarding RE for ML-enabled systems obtained from our previous study, an international survey [3]. We understand that bridging the gap between theory and practice is essential for RE maturity in such systems.

## 3 RESEARCH METHOD

### 3.1 Goal and Research Questions

This paper aims to characterize the current practices, perceptions, and challenges regarding RE for ML-enabled system projects in the Brazilian industry. From this goal, we established the following research questions:

- **RQ1. What are the contemporary practices adopted in Brazil regarding RE for ML-enabled systems?** This question aims to reveal how practitioners are currently approaching RE for ML in Brazilian companies by identifying trends, main methods, and the extent to which the degree of alignment with established industry practices. We refined *RQ1* into the following questions:
  - RQ1.1 Who is addressing the requirements of Brazilian ML-enabled system projects?
  - RQ1.2 How are requirements typically elicited in Brazilian ML-enabled system projects?
  - RQ1.3 How are requirements typically documented in Brazilian ML-enabled system projects?
  - RQ1.4 Which NFRs do typically play a major role in Brazilian ML-enabled system projects?
  - RQ1.5 Which activities are considered to be most difficult when defining requirements for Brazilian ML-enabled system projects?
- **RQ2. What are the main RE-related challenges faced by Brazilian practitioners working on ML-enabled system**

**projects?** Identifying these challenges in Brazil reflects the current maturity of these systems in the country. At the same time, it also informs the development of strategies to mitigate difficulties, helping to steer future research on the topic in a problem-driven manner. For this research question, we applied open and axial coding procedures to allow the problems to emerge from the open-text responses provided by the practitioners.

## 3.2 Survey Design

We extracted and analyzed data from our previous study, which presented an international survey [3, 24] that was conducted based on best practices of survey research [46], carefully conducting the steps below:

- **Step 1. Initial Survey Design.** We conducted a literature review on RE for ML [42] and combined our findings with previous results on RE problems [11] and the RE status quo [45] to provide the theoretical foundations for questions and answer options. Therefrom, the initial survey was drafted by software engineering and machine learning researchers of PUC-Rio (Brazil) with experience in R&D projects involving ML-enabled systems.

- **Step 2. Survey Design Review.** The survey was reviewed and adjusted based on online discussions and annotated feedback from software engineering and machine learning researchers of BTH (Sweden). Thereafter, the survey was also reviewed by the other co-authors.

- **Step 3. Pilot Face Validity Evaluation.** This evaluation involves a lightweight review by randomly chosen respondents. It was conducted with 18 Ph.D. students taking a Survey Research Methods course. They were asked to provide feedback on the clearness of the questions and to record their response time. This phase resulted in minor adjustments related to usability aspects and unclear wording. The answers were discarded before launching the survey.

- **Step 4. Pilot Content Validity Evaluation.** This evaluation involves subject experts from the target population. Therefore, we selected five experienced data scientists developing ML-enabled systems, asked them to answer the survey, and gathered their feedback. The participants had no difficulties answering the survey, which took an average of 20 minutes. After this step, the survey was considered ready to be launched.

The final survey started with a consent form describing the purpose of the study and stating that it was conducted anonymously. The remainder was divided into 15 demographic questions (D1 to D15) and three specific parts with 17 substantive questions (Q1 to Q17): seven on the ML life cycle and problems, five on requirements, and five on deployment and monitoring. This paper focuses on the demographics, the Problem Understanding and Requirements stage of the ML life cycle, and specific questions regarding requirements. Excerpts of the substantive questions related to this paper are shown in Table 1. The survey was implemented using the Unipark Enterprise Feedback Suite [1].

---

[1] https://www.unipark.com/en/survey-software/

**Table 1: Research questions and survey questions**

| RQ | Survey No. | Description | Type |
|---|---|---|---|
| - | ... | ... | ... |
| RQ2 | Q4 | According to your personal experience, please outline the main problems or difficulties (up to three) faced during the Problem Understanding and Requirements ML life cycle stage. | Open |
| - | ... | ... | ... |
| RQ1.1 | Q8 | Who is actively addressing the requirements of ML-enabled system projects in your organization? | Closed (MC) |
| RQ1.2 | Q9 | How were requirements typically elicited in the ML-enabled system projects you participated in? | Closed (MC) |
| RQ1.3 | Q10 | How were requirements typically documented in the ML-enabled system projects you participated in? | Closed (MC) |
| RQ1.4 | Q11 | Which Non-Functional Requirements (NFRs) typically play a major role in terms of criticality in the ML-enabled system projects you participated in? | Closed (MC) |
| RQ1.5 | Q12 | Based on your experience, what activities do you consider most difficult when defining requirements for ML-enabled systems? | Closed (MC) |
| - | ... | ... | ... |
| - | - | MC = Multiple Choice | - |

## 3.3 Data Collection

Our target population concerns professionals involved in building ML-enabled systems, including different activities, such as management, design, and development. Therefore, it includes practitioners in positions such as project leaders, requirements engineers, data scientists, and developers. We used convenience sampling, sending the survey link to professionals active in our partner companies, and also distributed it openly on social media.

In this paper, we excluded participants who informed on the survey that they had no experience with ML-enabled system projects and those working in other countries except Brazil. Data collection was open from January 2022 to April 2022. We received responses from 276 professionals; 188 completed all four sections of the survey, and of these, **72** were working in Brazil, which constituted our total sample. The average time to complete the survey was 20 minutes. We conservatively considered only the 72 fully completed survey responses from professionals working in Brazil.

## 3.4 Data Analysis Procedures

For data analysis purposes, given that all questions were optional, the number of responses varies across the survey questions. Therefore, we explicitly indicate the number of responses when analyzing each question.

Research questions *RQ1.1 - RQ1.5* concern closed questions, so we decided to use inferential statistics to analyze them. Our population has an unknown theoretical distribution (*i.e.*, the distribution of ML-enabled system professionals is unknown). In such cases, resampling methods - like bootstrapping - have been reported to be more reliable and accurate than inference statistics from samples [32, 46]. Hence, we use bootstrapping to calculate confidence intervals for our results, similar as done in [45]. In short, bootstrapping involves repeatedly taking samples with replacements and then calculating the statistics based on these samples. For each question, we take the sample of $n$ responses for that question and bootstrap $S$ resample (with replacements) of the same size $n$. We assume $n$ as the total valid answers of each question [10], and we set 1000 for $S$, which is a value that is reported to allow meaningful statistics [28]. Figure 1 summarizes the adopted bootstrapping method.

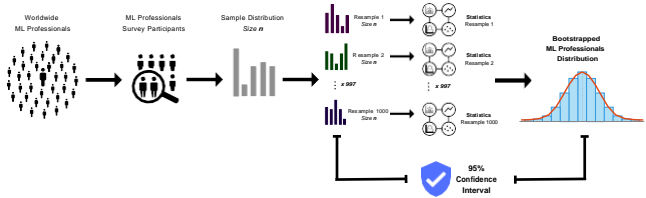

**Figure 1: Bootstrapping technique**

For research question *RQ2*, which seeks to identify the main challenges faced by practitioners involved in engineering ML-enabled systems related to problem understanding and requirements, the corresponding survey question is designed to be open text. We conducted a qualitative analysis using open and axial coding procedures from grounded theory [41] to allow the challenges to emerge from the open-text responses reflecting the experience of the practitioners. The primary author performed the qualitative coding procedures and subsequently reviewed them with the secondary author. Additionally, three researchers from academic and industry partners reviewed the resulting codes independently.

The questionnaire, the collected data, and the quantitative and qualitative data analysis artifacts, including Python scripts for the bootstrapping statistics, charts, and peer-reviewed qualitative coding spreadsheets, are available in our open science repository [38].

## 4 RESULTS

### 4.1 Study Population.

We focus specifically on the data obtained from Brazil as part of our previous study, which provided a larger international survey on ML-enabled systems engineering [24]. The study population consisted of 72 practitioners involved in data-driven projects across various industries in Brazil. These respondents held various roles, backgrounds, and professional experience. This diverse group provides a comprehensive view of the current practices, perceptions, and challenges related to RE for ML within the Brazilian context.

Figure 2 provides insights into the characteristics of the Brazilian participants involved in the survey. Regarding company size, the majority of participants (58.3%) are employed by companies with over 2000 employees and only 11.2% of them are employed by small companies as presented Figure 2 (a). In Figure 2 (b), we present participants' main roles. Data Scientists, Business Analysts, and Project Leads/Project Managers are the most common roles represented. Notably, the less assessed positions were Tester and Requirements Engineer, with one professional each.

Regarding ML-enabled system experience, in Figure 2 (c), most participants reported having 1 to 2 years of working experience. Closely, another significant portion of respondents indicated a higher experience range of 3 to 6 years. This proportion emphasizes a balanced population of beginner and experienced professionals. It is noteworthy that regarding participants' educational background,

87.5% mentioned having a bachelor's degree in computer science, information systems, statistics, or electrical engineering. Moreover, 45.83% held master's degrees in computer science, data science, or electrical engineering. Lastly, 19.44% completed Ph.D. programs in computer science, physics, or computer engineering.

### 4.2 Problem Understanding and Requirements ML Life Cycle Stage

In the survey, based on the nine ML life cycle stages presented by Amershi *et al.* [4] and the CRISP-DM industry-independent process model phases [39], we abstracted seven generic life cycle stages [21] and asked about their perceived relevance and difficulty. The answers presented in Figure 3 and 4 revealed that ML practitioners are extremely worried about requirements given that the *Problem Understanding and Requirements* stage is clearly perceived as the most relevant and most complex life cycle stage.

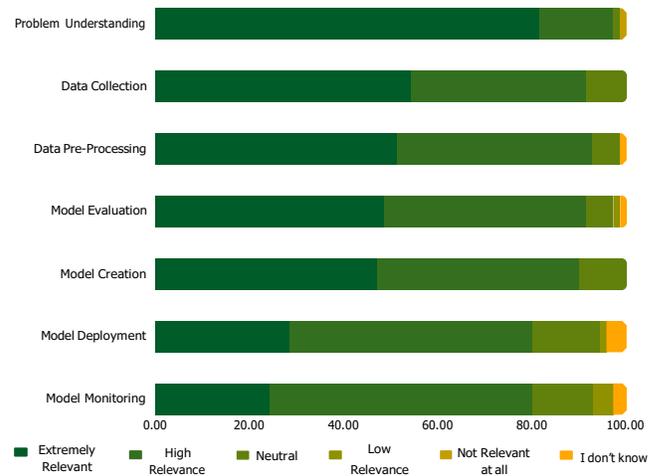

**Figure 3: Perceived Relevance of each ML life cycle stage**

### 4.3 Contemporary RE practices for ML-enabled Systems

*4.3.1 [RQ1.1] Who is addressing the requirements of ML-enabled system projects?* The proportion of positions reported to address the requirements of ML-enabled system projects in the bootstrapped samples is shown in Figure 5 together with the 95% confidence interval. The $N$ in each figure caption is the number of participants that answered this question. We report the proportion $P$ of the participants that checked the corresponding answer and its 95% confidence interval in square brackets.

It is possible to observe that Data Scientists were most associated with requirements in ML-enabled systems with **P = 61.389 [60.955, 61.822]**, followed by Project Leaders (**P = 49.6 [49.219, 49.981]**), Business Analysts (**P = 28.339 [28.024, 28.653]**), and Developers (**P = 21.386 [21.061, 21.71]**). The less associated roles within requirements addressing were Solution Architects (**P = 11.563 [11.353, 11.773]**), Requirements Engineers (**P = 8.46 [8.281, 8.639]**), and Testers (**P = 1.481 [1.397, 1.566]**). Several isolated options were

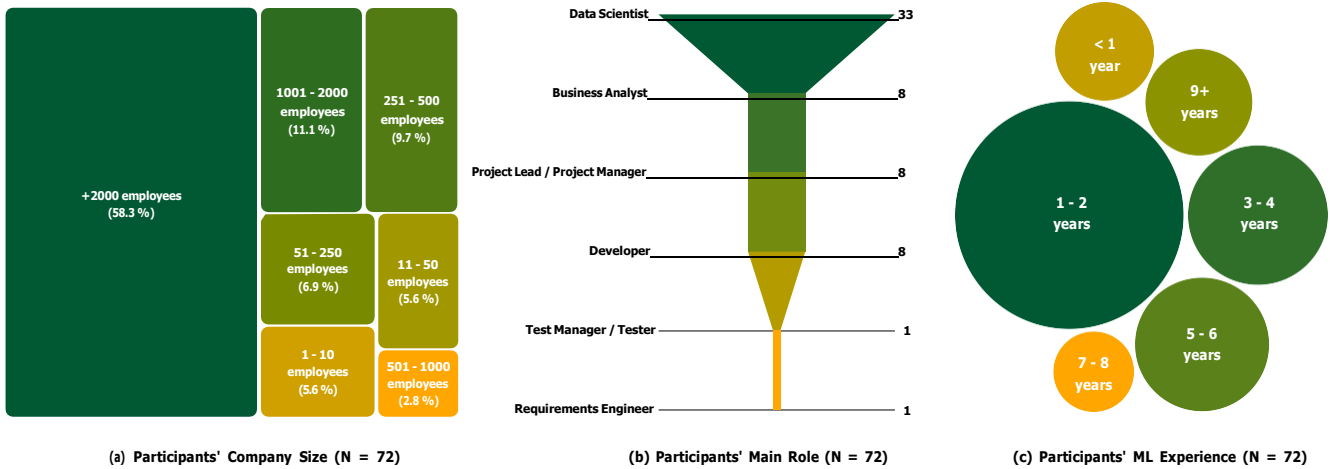

(a) Participants' Company Size (N = 72)

(b) Participants' Main Role (N = 72)

(c) Participants' ML Experience (N = 72)

**Figure 2: Practitioners' demographics: company size, roles, and ML experience**

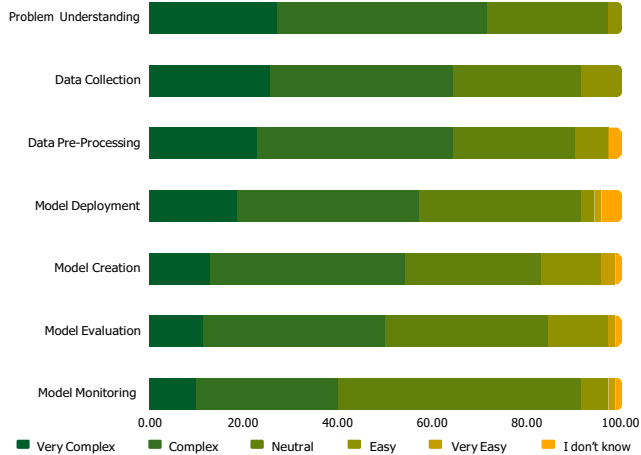

**Figure 4: Perceived Difficulty of each ML life cycle stage**

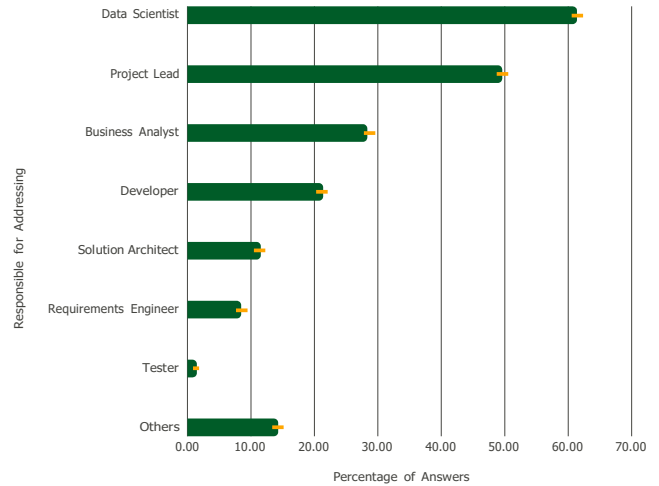

**Figure 5: Roles addressing requirements of ML-enabled systems (N = 70)**

mentioned in the "Others" field (*e.g.*, Machine Learning Engineer, Data Lead, and Tech Lead), altogether summing up 14% and not significantly influencing the overall distribution (**P = 14.303 [14.032, 14.573]**).

### 4.3.2 [RQ1.2] How are requirements typically elicited in ML-enabled system projects?

As presented in Figure 6, practitioners reported interviews as the most commonly used technique (**P = 69.399 [69.062, 69.735]**), followed (or complemented) by workshops (**P = 47.296 [46.958, 47.634]**), prototyping (**P = 41.638 [41.292, 41.983]**), and scenarios (**P = 40.221 [39.841, 40.6]**). The least used elicitation technique was observation, with **P = 35.896 [35.535, 36.257]**. In the "Others" field, the Objective and Key Results (OKRs) system and informal meetings were mentioned, but with a much lower proportion (**P = 8.357 [8.156, 8.558]**).

### 4.3.3 [RQ1.3] How are requirements typically documented in the ML-enabled system projects?

Figure 7 shows Notebooks as the most frequently used documentation format with **P = 46.504 [46.129, 46.879]**, followed by User Stories (**P = 30.715 [30.374, 31.057]**), Vision Documents (**P = 21.304 [21.008, 21.6]**), Prototypes (**P = 21.182 [20.895, 21.468]**), Requirements Lists (**P = 19.713 [19.431, 19.994]**), and Data Models (**P = 19.669 [19.352, 19.986]**). Surprisingly, almost 17% mentioned that requirements are not documented at all with **P = 16.917 [16.632, 17.201]**. Some isolated options were mentioned in the "Others" field (*e.g.*, Notion, Github, and Confluence) with **P = 12.668 [12.429, 12.906]**.

### 4.3.4 [RQ1.4] Which Non-Functional Requirements (NFRs) do typically play a major role in terms of criticality in the ML-enabled system projects?

Regarding NFRs (Figure 8), practitioners show a significant concern with some ML-related NFRs, such as data quality (**P = 69.103 [68.75, 69.456]**), model explainability (**P = 37.825 [37.464, 38.187]**), and model reliability (**P = 36.721 [36.341, 37.101]**). Some

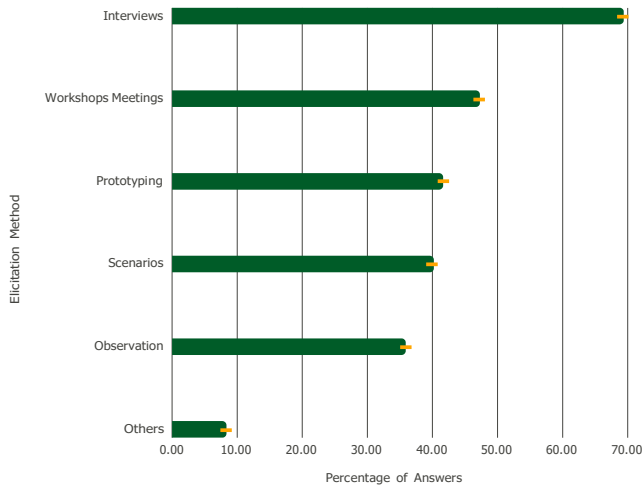

**Figure 6: Requirements Elicitation techniques of ML-enabled systems (N = 72)**

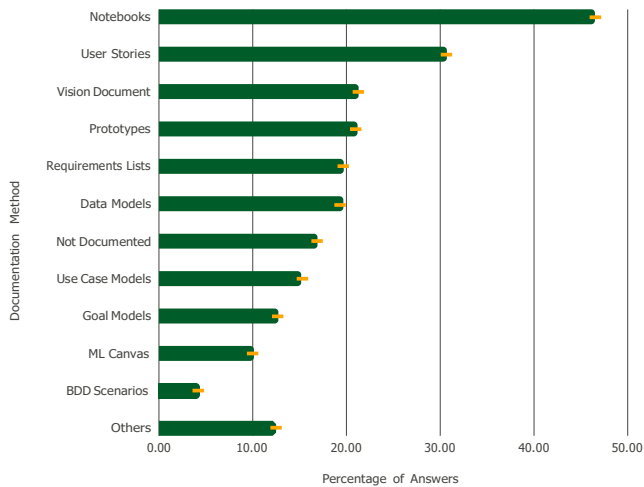

**Figure 7: Requirements Documentation of ML-enabled systems (N = 71)**

NFRs regarding the whole system were also considered important, such as system performance (**P = 35.2 [34.874, 35.526]**), system maintainability (**P = 25.441 [25.122, 25.76]**), and system usability (**P = 25.175 [24.828, 25.521]**). A significant number of participants informed that NFRs were not considered within their ML-enabled system projects (**P = 12.623 [12.376, 12.869]**).

*4.3.5 [RQ1.5] Which activities are considered most difficult when defining requirements for ML-enabled systems?* The answer options to this question were based on the literature regarding requirements [45] and requirements for ML [42]. Furthermore, we left the "Others" option to allow new activities to be added, but nothing new was informed. In this context, we show in Figure 9 that the respondents considered managing customer expectations is the most difficult task (**P = 71.554 [71.191, 71.916]**), followed by aligning

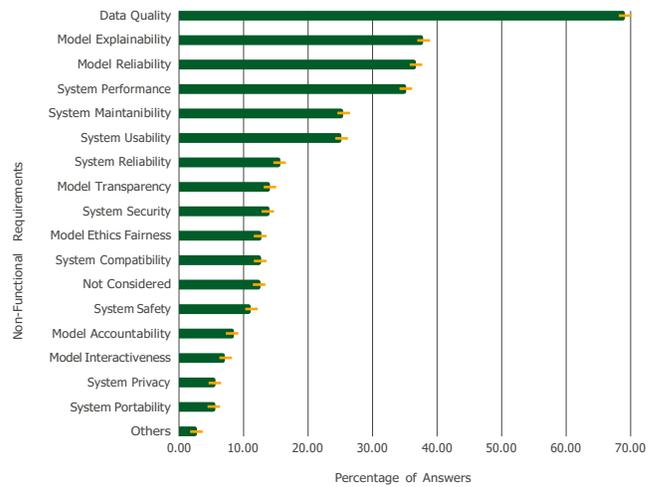

**Figure 8: Critical Non-Functional Requirements of ML-enabled systems (N = 71)**

requirements with data (**P = 53.556 [53.197, 53.915]**), resolving conflicts (**P = 42.346 [41.987, 42.706]**), managing changing requirements (**P = 40.915 [40.574, 41.257]**), selecting metrics (**P = 32.079 [31.738, 32.42]**), and elicitation and analysis task (**P = 26.72 [26.418, 27.021]**).

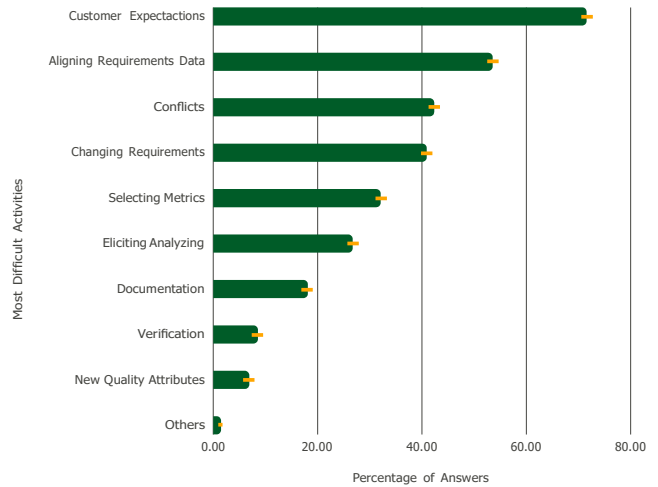

**Figure 9: Most Difficult RE activities in ML-enabled systems (N = 71)**

## 4.4 Main RE-related challenges in ML-enabled System Projects

Regarding the main concerns during each ML life cycle stage, we asked participants to inform up to three challenges related to each ML life cycle stage in an open-text answer. The main challenges related to the Problem Understanding and Requirements stage emerged from open coding applied to all of the 109 open-text answers provided for this stage.

We incorporated axial coding procedures to provide an easily understandable overview, relating the emerging codes to categories. We started with the sub-categories *Input*, *Method*, *Organization*, *People*, and *Tools*, as suggested for problems in previous work on defect causal analysis [20]. Based on the collected data, we merged the *Input* and *People* categories, as it was difficult to separate between the two, given the concise and short answers provided by the participants. We also renamed the *Tools* category into *Infrastructure* and identified the need to add a new category related to *Data*. It is noteworthy that these categories were identified considering the overall coding for the seven ML life cycle stages. At the same time, this paper focuses on the Problem Understanding and Requirements stage.

In Figure 10, we present an overview of the resulting codes' frequencies using a probabilistic cause-effect diagram (*e.g.*, fishbone diagram), which provides a comprehensive overview and it was introduced for causal analysis purposes in previous work [22, 23]. The percentages are just frequencies of occurrence of the codes, then the sum of all code frequencies is 100%. It is important to notice that the highest frequencies within each category are organized closer to the middle.

It is possible to observe that most of the reported challenges are related to the *Input* category, followed by *Method* and *Organization*. Within the *Input* category, the main challenges report difficulties in understanding the problem, the business domain, and unclear goals and requirements. In the *Method* category, the prevailing reported challenges concern difficulties in managing expectations, experienced data science knowledge, and establishing effective communication. Finally, in the *Organization* category, the lack of customer or domain expert availability and engagement and the lack of time dedicated to requirements-related activities were mentioned. Our summary focuses on the most frequently mentioned challenges, although less frequent ones may still be relevant in practice. For instance, computational constraints or lack of data quality and pre-processing can directly affect ML-related possibilities and requirements.

## 5 DISCUSSION

Our previous results reflected an international perspective regarding RE for ML-enabled systems [3]. Given the importance of Brazil in this previous study and the growing interest of Brazilian companies in terms of ML, we bring a deeper and focused analysis of Brazil's practices, problems, and perceptions in this paper. For instance, we have an intriguing distribution of roles that address requirements. Contrary to conventional expectations, where either requirements engineers or business analysts [47] could be in charge of requirements, we have data scientists taking the rein. Unlike our previous findings [3], we have data scientists as the majority in RE addressing activity in Brazil, which reassures their importance beyond coding in companies [7, 26]. The nature of ML-enabled systems is based on data-driven insights, which may explain the importance of addressing activity in this role. However, lacking well-established methods and practices in this domain may lead to project failure [11].

Regarding elicitation techniques, our survey revealed again that practitioners don't escape from traditional requirements elicitation techniques (interviews, prototyping, scenarios, workshops, and observation), even with a free-text option available. Unlike our previous results where Workshops were less used [3], in Brazil, our results for the elicitation techniques are comparable for traditional RE [45]. This could be related to the fact that most practitioners work in large companies, which typically have professionals experienced in conducting such workshops for traditional software systems and have now extended these practices to ML-enabled systems.

In terms of requirements documentation, computational notebooks, which are interactive programming environments that can be used to process data and create ML models, appear, as reported previously [3], as the most used tool for documenting requirements. Its rapid way of producing and generating code turned notebooks into an important tool for data scientists; however, like a hammer, it could be misused [37] as a symptom of the lack of awareness of RE specification practices and tools. Moreover, a proportion of almost 16% mentioned that requirements were not documented at all, which may cause overall software project failure [11]. In Brazil, we have reported that Vision Documents are more prominent than in other parts of the world, and despite being closely related to Prototypes, our previous finding had Requirements Lists as the third most used method, and now it appears as the sixth option. ML Canvas, which was designed to tackle this activity, is one of the least used methods, along with BDD Scenarios.

With respect to NFRs, there are slight differences between how worldwide practitioners face NFR [3] and how Brazilians do. In general, the most considered concerns are ML-related NFR, such as data quality, model reliability, and model explainability, as previously reported in [15, 44]. However, we also observed system-related concerns like system performance, usability, and maintainability. In conventional software systems, there are several negative impacts of missing NFRs on software-related projects [11]. However, again, the same proportion of practitioners (more than 10%) do not even consider NFRs in their ML-enabled system projects, which can be seen as another indicator of the lack of overall attention to the importance of RE in the industrial ML-enabled systems engineering context.

The survey also revealed the most difficult activities perceived by practitioners in Brazil when defining requirements for ML-enabled systems. The difficulties reported by Brazilian practitioners are comparable to the ones reported previously [3] and with previous literature, but now it appears in a wider industrial scope. Managing customer expectations [18], aligning requirements with data [35, 42], changing requirements [25], and selecting proper metrics [44] were previously reported as difficulties, which emphasizes the importance of effective communication and technical expertise to bridge the gap between aspirations and technological feasibility.

Finally, we contributed to the RE-related problems faced by practitioners in ML-enabled system projects in Brazil, which are slightly different from our previous study [3]. Still, the main issues relate to difficulties in problem and business understanding, managing expectations, and low customer/domain expert availability/engagement. These issues have comparable counterparts in the conventional RE problems [11]. Table 2 shows the strong relationship between problems in ML-enabled systems and traditional contexts. As comparable problems may have comparable solutions, adopting established RE practices (or adaptations of such practices) may help improve ML-enabled system engineering, but this will demand further empirical evaluations and is not the scope of the paper.

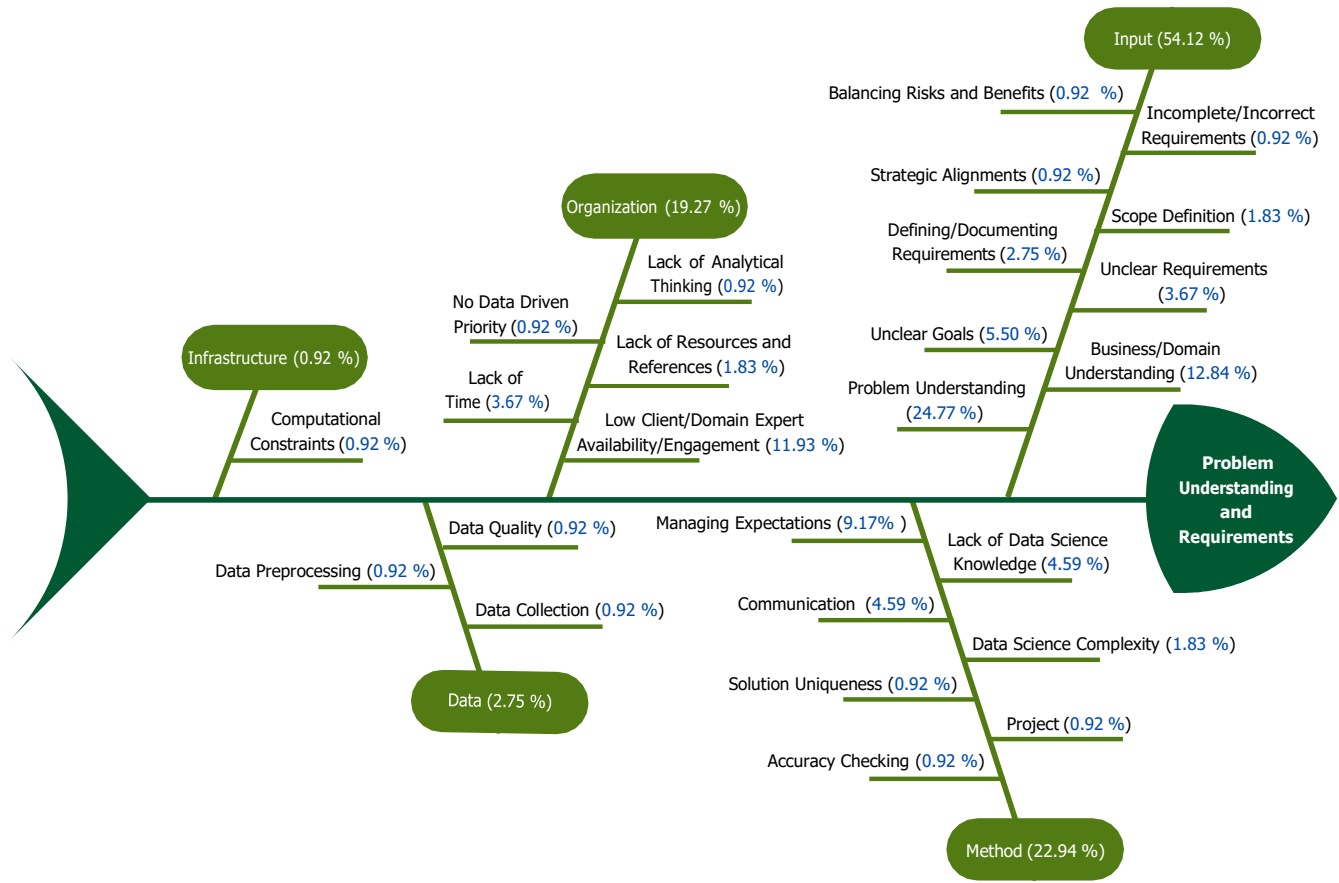

**Figure 10: Main Problems faced during Problem Understanding and Requirements**

## 6 THREATS TO VALIDITY

We identified some threats while planning, conducting, and analyzing the survey results. Henceforward, we list these potential threats organized by the survey validity types presented in [31].

**Face and Content Validity**. Face and content validity threats include bad instrumentation and inadequate explanation of constructs. To mitigate these threats, we involved several researchers in reviewing and evaluating the questionnaire regarding the format and formulation of the questions, piloting it with 18 Ph.D. students for face validity and five experienced data scientists for content validity.

**Criterion Validity**. Threats to criterion validity include not surveying the target population. We clarified the target population in the consent form (before starting the survey). We also considered only complete answers (*i.e.*, answers of participants that answered all four survey sections) and excluded participants that informed having no experience with ML-enabled system projects. Moreover, an important aspect is a possible bias in our results, given only one requirements engineer answered our survey. We explain that the explicit Requirements Engineer position is uncommon even in conventional software engineering contexts and that mainly other positions, like Business Analysts, are responsible for RE-related tasks [47]. Hence, we believe that not having many requirements engineers in our sample is expected and positive in terms of representativeness, and just reflects that they are typically not part of ML-enabled system projects.

**Construct Validity**. We ground our survey's questions and answer options on theoretical background from previous studies on RE [11, 45] and a literature review on RE for ML [42]. A threat to construct validity is inadequate measurement procedures and unreliable results. To mitigate this threat we follow recommended data collection and analysis procedures [46].

**Reliability**. One aspect of reliability is statistical generalizability. We could not construct a random sample that systematically covers all types of professionals involved in developing ML-enabled systems, as there is still no generalized understanding of what such a population looks like. Nevertheless, the experience and background profiles of the subjects are comparable to the profiles of ML teams, as shown in Microsoft's study [26]. We used bootstrapping to deal with the random sampling limitation and only employed confidence intervals, conservatively avoiding null hypothesis testing. Another reliability aspect concerns inter-observer reliability, which we improved by including independent peer review in all our qualitative analysis procedures and making all the data and analyses openly available online [38].

**Table 2: Comparison between problems on ML-enabled and traditional systems**

| Traditional RE Problem | ML RE Problem |
| --- | --- |
| Incomplete and/or hidden requirements | [Input] Incomplete/incorrect requirements |
| Communication flaws between project team and customer | [Method] Communication |
| Moving targets (changing goals, business processes, and/or requirements) | [Input] Unclear goals |
| Underspecified requirements that are too abstract | [Input] Unclear requirements |
| Timeboxing/Not enough time in general | [Organization] Lack of time |
| Stakeholders with difficulties in separating requirements from known solution designs | [Organization] Lack of analytical thinking |
| Insufficient support by customer | [Organization] Low client/domain expert availability/engagement |
| Weak access to customer needs and/or business information | [Organization] Lack of resources and references |

## 7 CONCLUSIONS

Literature suggests that RE can help to tackle challenges in ML-enabled system engineering [42]. Recent literature studies (*e.g.*, [1, 35, 42]) and industrial studies (*e.g.*, [7, 44]) on RE for ML-enabled systems have been important to help to understand the literature focus and industry needs. However, the study of industrial practices, perceptions, and challenges is still isolated and not yet representative.

We build upon prior research to enhance the empirical evidence on current practices, perceptions, and challenges in the field of RE for ML. This study analyzes a subset of data from our previous study, which presented an international survey [3], focusing on responses from 72 practitioners involved in the development of ML-enabled systems in Brazil. We applied bootstrapping with confidence intervals for quantitative statistical analysis and open and axial coding for qualitative analysis of RE challenges. The results reinforce the findings of previous studies [15, 44], emphasizing the importance of non-functional requirements, such as data quality, model reliability, and explainability. They also highlight challenges, including managing customer expectations and addressing ambiguities in requirements specifications [35, 42].

In addition, the analysis of data uncovered several new and noteworthy aspects. Notably, data scientists are increasingly leading RE activities in the development of ML-enabled systems, with interactive notebooks serving as a primary method for documenting requirements. The survey also highlighted several challenges faced by practitioners, such as difficulties in problem and business understanding, difficulties in managing expectations, unclear requirements, and lack of domain expert availability and engagement.

Overall, when comparing RE practices and challenges within ML-enabled systems with conventional RE practices [45] and challenges [11], we identified significant variations in the practices but comparable underlying problems. Proposing solutions for these problems is part of future research and is not in the scope of this paper, as it would demand proper empirical evaluations through different empirical strategies (e.g., action research, case studies, controlled experiments). However, we truly believe that comparable challenges may have comparable solutions. In this sense, we advocate for adapting and disseminating RE-related practices for engineering ML-enabled systems.

## ONLINE RESOURCES

Our data, artifacts, and additional resources are openly available at Zenodo [38] under the Creative Commons Attribution license.

## ACKNOWLEDGMENTS

We thank the Brazilian Council for Scientific and Technological Development (CNPq process #312275/2023-4) and StoneCo Ltd (research project "Software Engineering for Data Science and Artificial Intelligence") for financial support.